# Barbosa, Uniform Polynomial Time Bounds, and Promises[*]

Lane A. Hemaspaandra,[†] Kyle Murray, and Xiaoqing Tang
Department of Computer Science
University of Rochester
Rochester, NY 14627, USA

June 6, 2011

**Abstract**

This note is a commentary on, and critique of, André Luiz Barbosa's paper entitled "P != NP Proof." Despite its provocative title, what the paper is seeking to do is not to prove $P \neq NP$ in the standard sense in which that notation is used in the literature. Rather, Barbosa is (and is aware that he is) arguing that a different meaning should be associated with the notation $P \neq NP$, and he claims to prove the truth of the statement $P \neq NP$ in his quite different sense of that statement. However, we note that (1) the paper fails even on its own terms, as due to a uniformity problem, the paper's proof does not establish, even in its unusual sense of the notation, that $P \neq NP$; and (2) what the paper means by the claim $P \neq NP$ in fact implies that $P \neq NP$ holds even under the standard meaning that that notation has in the literature (and so it is exceedingly unlikely that Barbosa's proof can be fixed any time soon).

## 1 Introduction

This note is a brief commentary on an `arXiv.org` article by Barbosa that has the provocative title "P != NP Proof."

Despite the title, the paper is not really purporting to resolve the P versus NP problem in the standard sense. Rather, the author redefines the notation $P \neq NP$, casting it in effect to be a statement that there exists at least one (potentially restricted, promise-like) domain on which P and NP differ. Barbosa is aware of this distinction and openly discusses the fact, saying "I have restated the P versus NP question on my own terms."

What we observe in this note is that the paper fails even on its own terms: Due to a uniformity problem, it does not prove what it claims to. But can the proof somehow be fixed? We note that what Barbosa means by $P \neq NP$ in fact (if true) itself implies (in the standard sense) $P \neq NP$. And so, despite the fact that Barbosa's paper is at the moment we are writing this on its 38th public version in less than two years, we feel it is unlikely that Barbosa's altered notion of what "P $\neq$ NP" means can be established any time soon.

---

[*]Also appears as URCS-TR-2011-969.
[†]URL: `www.cs.rochester.edu/u/lane`. Supported in part by grant NSF-CCF-0915792 and a Friedrich Wilhelm Bessel Research Award.

## 2  Context-Setting

This entire note is based on the most recent version of Barbosa's paper available as this note was being drafted, namely, Version 38 (the version of May 21, 2011) of arXiv.org report 0907.3965 [Bar11].

We write this simply as a very brief note, so we assume that the reader has Barbosa's paper in hand, and we don't even set the stage with his definitions (except in Section 3 we briefly deal with some clarifications that are needed).

Readers may wonder whether it makes sense to respond to papers that have P = NP or P $\neq$ NP in their title. In the particular case of the present critique, the critique grew out of a teaching exercise in an undergraduate course, in particular, a problem-solving course that has a tradition of, as an exercise (and arguably as a small service to the literature regarding clarifying the meaning and correctness of claims that have been publicly made), looking at papers that might seem to claim to resolve P-versus-NP or the complexity of graph isomorphism, in order to understand what the papers are claiming, and whether they achieve their claims, and to share what is learned from that. The present paper is authored, jointly with the course's instructor, by the two students who looked at Barbosa's paper. (Among critiques to come out of earlier years' instances of the same course are the papers of Sabo, Schmitt, and Silverman [SSS07], Clingerman, Hemphill, and Proscia [CHP08], Christopher, Huo, and Jacobs [CHJ08], and Ferraro, Hall, and Wood [FHW09], which are respectively responding to extremely strong P-versus-NP claims made by Feinstein, Yatsenko, Gubin, and Aslam. We commend to the reader's attention Woeginger's valuable, excellent, quite remarkably web page on attempted resolutions/refutations regarding P versus NP [Woe].

## 3  Critique

Before beginning the critique, we must cover some housekeeping items. We won't repeat Barbosa's definitions here, but we do need a notation that will let us distinguish between his notions and the standard notions, and we also need to smooth out some of the arguable confusions in his paper, so we can meaningfully critique what he is seemingly trying to express. Briefly, he is proposing a promise-like notion in which instead of having $\Sigma^*$ as the domain, the domain is instead (somewhat magically) restricted to some set $L_z$ ($z$ is not any particular item here; the notation $L_z$ is simply a notation from Barbosa), $L_z \subseteq \Sigma^*$. This occurs in Definition 3.5 of [Bar11]. Barbosa does not require that $L_z$ be, for example, a recursive set. It can be any set. To avoid any confusion, let us call the version of NP created (for the restriction $L_z$) by Barbosa's definition NP[$L_z$]. And Barbosa's paper's notion of what he proposes as a good, new, changed semantics for the notation P $\neq$ NP is in fact what one would more naturally denote by the claim

$$(\exists L_z \subseteq \Sigma^*)[P[L_z] \neq NP[L_z]].$$

In a more common way of speaking of such things, one might express this as follows (and Barbosa as his Section 3.3.2 does mention that his notion is essentially the same as the notion of promise problems [ESY84,Sel88]): There exists a promise problem that has a solution in NP but not in P.[1] (This already should make clear to the reader the seemingly insurmountable problems one would face to prove Barbosa's claim, but let us first clear up the definition before returning to that later.) As to Barbosa's definition (still Definition 3.5 of his paper), we mention in passing that (a) the big-O isn't used properly (rather, he needs to before the universal quantification on $x$ fix a polynomial bounding the length of the certificates; we from here on assume that his definition is viewed as being modified to do that); and (b) he defines $NP[L_z]$—which he extremely unfortunately denotes simply NP—with $L_z$ simply appearing in the definition without any clear quantification, and so formally it isn't clear that his notion of $P \neq NP$ even manages to link the P and the NP to be based on the same domain $L_z$ (but having them be the same $L_z$ is the clear intent of Barbosa's paper, and we from here on assume that that is the case, i.e., we treat Barbosa as if he is trying to prove $(\exists L_z \subseteq \Sigma^*)[P[L_z] \neq NP[L_z]]$, since that is indeed what his paper is (seeking to) prove).

And with our apologies for having taken so long to get to such brief, straightforward points, we can now make our two comments about the content of Barbosa's paper.

Our first point is that regarding the test set ("XG-SAT") that he claims to prove, for a particular and rather complex and clever set $L_z$ that he defines, is in $NP[L_z]$ but is not in $P[L_z]$, his proof that his test set belongs to $NP[L_z]$ seems flawed. In particular, although it is called "XG-SAT," the set is actually about taking codes of programs and running them. The proof attempts to (in the complicated and arguably ambiguous or ill-defined "Definition 2.1") as a promise require each machine to (mostly) run in polynomial time. However, there is no single, shared polynomial, and Barbosa goes out of his way to state that he is not padding things down (in the way that routinely is done when making universal complete sets, for example) with a padding string whose length is the allowed running time. The problem with all this is that some machines will run in linear time, some will run in quadratic time, some in cubic time, and so on, and so the set XG-SAT has no obvious single polynomial upper-bounding the nondeterministic running time of a NTM accepting it. So the proof that XG-SAT belongs to NP[$L_z$] seems invalid.

---

[1]One might worry that Barbosa's definition (see his paper) is not really clear about what it means when it says that the P-time predicate defining membership in $NP[L_z]$ is P-time—whether it means that the underlying machine is in P-time over all inputs that belong to $L_z$ or whether it is P-time globally (over $\Sigma^*$). However, there is no issue here, as if the former holds, one can take a polynomial $p'$ of the form $n^i + i$ that majorizes that machine's polynomial run-time bound on $L_z$, and can then take one's original underlying machine $M$ and from it build a new machine, $M'$, that on each input simulates $M$ for a number of steps that is $p'$ applied to the input length. Note that $M'$ is globally P-time. For this reason, we from here on will assume that languages in $NP[L_z]$ and $P[L_z]$ are defined and instantiated by machines that globally obey nondeterministic or deterministic polynomial time bounds. For completeness, we mention in passing that Barbosa actually in the stated definition defines the P-time predicate R's domain as being $L_z \times \Sigma^*$, and so he may indeed be thinking of the time bound as applying just on that region.

Now we turn to our second point. Can one hope to somehow fix Barbosa's proof and establish his claim? Recall that his claim is that

$$(\exists L_z \subseteq \Sigma^*)[P[L_z] \neq NP[L_z]].$$

But suppose that that holds. That means (keeping in mind the comments of Footnote 1 about slapping on a global clock as needed) that there is an NPTM, call it $N'$, whose language when restricted to $L_z$ differs, for every DPTM $M'$, from the restriction to $L_z$ of the language of $M'$. However, note that if P = NP (in the standard sense the literature has for what that means), then the language of $N'$ is, since P = NP, accepted by some DPTM, call it $\widehat{M}$. And so certainly the restriction of $L(\widehat{M})$ to $L_z$ is identical to the restriction of $L(N')$ to $L_z$, thus putting this into P[$L_z$], contrary to what we assumed. Put another way, clearly it follows easily from the definitions that: If $(\exists L_z \subseteq \Sigma^*)[P[L_z] \neq NP[L_z]]$, then P $\neq$ NP. So proving Barbosa's main result would implicitly separate NP from P in the standard sense of the literature. Thus we find it unlikely that Barbosa's main result can be correctly proven in the immediate future (or at least it cannot be proven without the prover winning the Clay Institute's million-dollar prize).

**Acknowledgments** We thank the members of the Spring 2011 CSC200/200H Undergraduate Problem Seminar for a memorable course, and we thank course TA Jake Scheiber for his feedback on an earlier version of this paper.